\documentclass[runningheads]{llncs}
\usepackage{graphicx}
\usepackage{booktabs}
\usepackage{amssymb}
\begin{document}
\title{Segmentation of Multimodal Myocardial Images Using Shape-Transfer GAN}
	%\thanks{Supported by organization x.}}
%
%\titlerunning{Abbreviated paper title}
% If the paper title is too long for the running head, you can set
% an abbreviated paper title here
%
\author{Xumin Tao\inst{1,2} \and
	Hongrong Wei\inst{1,2} \and
	Wufeng Xue\inst{1,2}\thanks{corresponding author}
	\and Dong Ni\inst{1,2}}
\authorrunning{X. Tao, H. Wei, W. Xue, D. Ni}
% First names are abbreviated in the running head.
% If there are more than two authors, 'et al.' is used.
%
\institute{The National-Regional Key Technology Engineering Ultrasound, Guangdong Key Laboratory for Biomedical
	Measurements and Ultrasound Imaging, School of Biomedical Engineering, Health Science Center, Shenzhen
	University, Shenzhen, China \and
	Medical Ultrasound Image Computing (MUSIC) Lab\\
	\email{xwolfs@hotmail.com}}
\maketitle              % typeset the header of the contribution
\begin{abstract}
Myocardium segmentation of late gadolinium enhancement (LGE) Cardiac MR images is important for evaluation of infarction regions in clinical practice. The pathological myocardium in LGE images presents distinctive brightness and textures compared with the healthy tissues, making it much more challenging to be segment. Instead, the balanced-Steady State Free Precession (bSSFP) cine images show clearly boundaries and can be easily segmented. Given this fact, we propose a novel shape-transfer GAN for LGE images, which can 1) learn to generate realistic LGE images from bSSFP with the anatomical shape preserved, and 2) learn to segment the myocardium of LGE images from these generated images. It’s worth to note that no segmentation label of the LGE images is used during this procedure. We test our model on dataset from the Multi-sequence Cardiac MR Segmentation Challenge. The results show that the proposed Shape-Transfer GAN can achieve accurate myocardium masks of LGE images.

\keywords{Segmentation \and LGE \and Cross-modality \and Shape Transfer.}
\end{abstract}
\section{Introduction}
Late gadolinium enhancement (LGE) MRI technology can accurately identify myocardial infarction(MI), myocardial fibrosis and cardiac amyloid and other diseases. Its good spatial resolution and tissue specificity have unique advantages in the diagnosis of various types of myocardial lesions. To this end, correct segmentation of LGE CMR images is a prerequisite of quantitative evaluation.

While recent advancements in deep neural network have results in many accurate models of automatic segmentation of cardiac left/right ventricle (LV/RV) from bSSFP cine images, only a few efforts have been given to segmentation of cardiac structures from LGE images. Contrary to bSSFP cine image where the myocardium and the background blood pool have different intensity distributions and can be well discriminated, the intensity of LGE images is heterogeneous for the myocardium and the boundary of the pathological part is even invisible. 

Recently proposed methods of LGE segmentation include model-based~\cite{ref_article1} and learning-based ones~\cite{ref_article2,ref_article3}. Zhuang et al. (2018) used multivariate mixture model to describe the likelihood of multi-source images in a common space and model the motion shift of different slices with a rigid transformation. After iteratively registration and segmentation, the model achieved good myocardial segmentation. However, the complexity of the model may hinder it from effective application in practice~\cite{ref_article2}. Xiong et al. (2019) proposed a dual fully convolutional neural network to extract global and local structures from MRI slices of different resolutions for 3D left atrium segmentation from LGE images. The network was trained with a dataset of 154 subjects and achieved accurate segmentation results~\cite{ref_article3}. Yue et al. (2019) used a deep neural network SRSCN, which incorporated shape prior and slice spatial information as regularization for LGE cardiac segmentation~\cite{ref_article1}. After being trained with LGE images of 25 patients, it can segment the LV, myocardium, and RV well. A drawback of these learning-based methods is that they require large manually labeled LGE images for model training, which is not always available and more prone to errors or an accurate registration between the cine MRI and LGE MRI. 

The MS-CMRSeg 2019 challenge that held in conjunction with STACOM at MICCAI 2019 provides an open and fair platform for the multi-sequence ventricle and myocardium segmentation. However, there are only LGE images of 5 patients with ground truth label for training. This adds more difficulty during the development of learning-based model besides the above-mentioned ones. To relieve the problems of insufficient training labels, we proposed to generate plenty of image-label pairs by generative adversarial network (GAN). Goodfellow et al.(2016) first proposed GAN and achieved impressive results in generating realistic images from noisy input vectors~\cite{ref_article4}. Various strategies have been devoted to the development of GAN to improve the quality of the generated fake images~\cite{ref_article7,ref_article8} or to learn the disentangled representations that are aware of high-level semantic context. For our work, high quality of generated image-label pair is of critical importance to the final performance. To this end, we make use of the recently proposed CycleGAN~\cite{ref_article5}, which employed a cycled reconstruction loss to ensure the consistency between the input and output domains.
 
We propose a novel method, shape-transfer GAN, for the segmentation of LGE cardiac images, without ground truth labels. Specifically, we introduce a shape preservation term to make the generated LGE images share the same myocardium shape with that of the input bSSFP image. In such a way, the proposed shape-transfer GAN is capable of generating realistic LGE images, and in the meantime learning how to segment these generated images. Without labels of real LGE images for finetuning, the obtained segmentor can be directly applied for segmentation of real LGE images. The method obtains good performance on LGE images of 40 patients, with dice metric of 0.847, 0.776, 0.686 for LV, RV and myocardium, respectively.

\section{Method}
\begin{figure}[!t]
	\centering
	\includegraphics[width=0.85\textwidth]{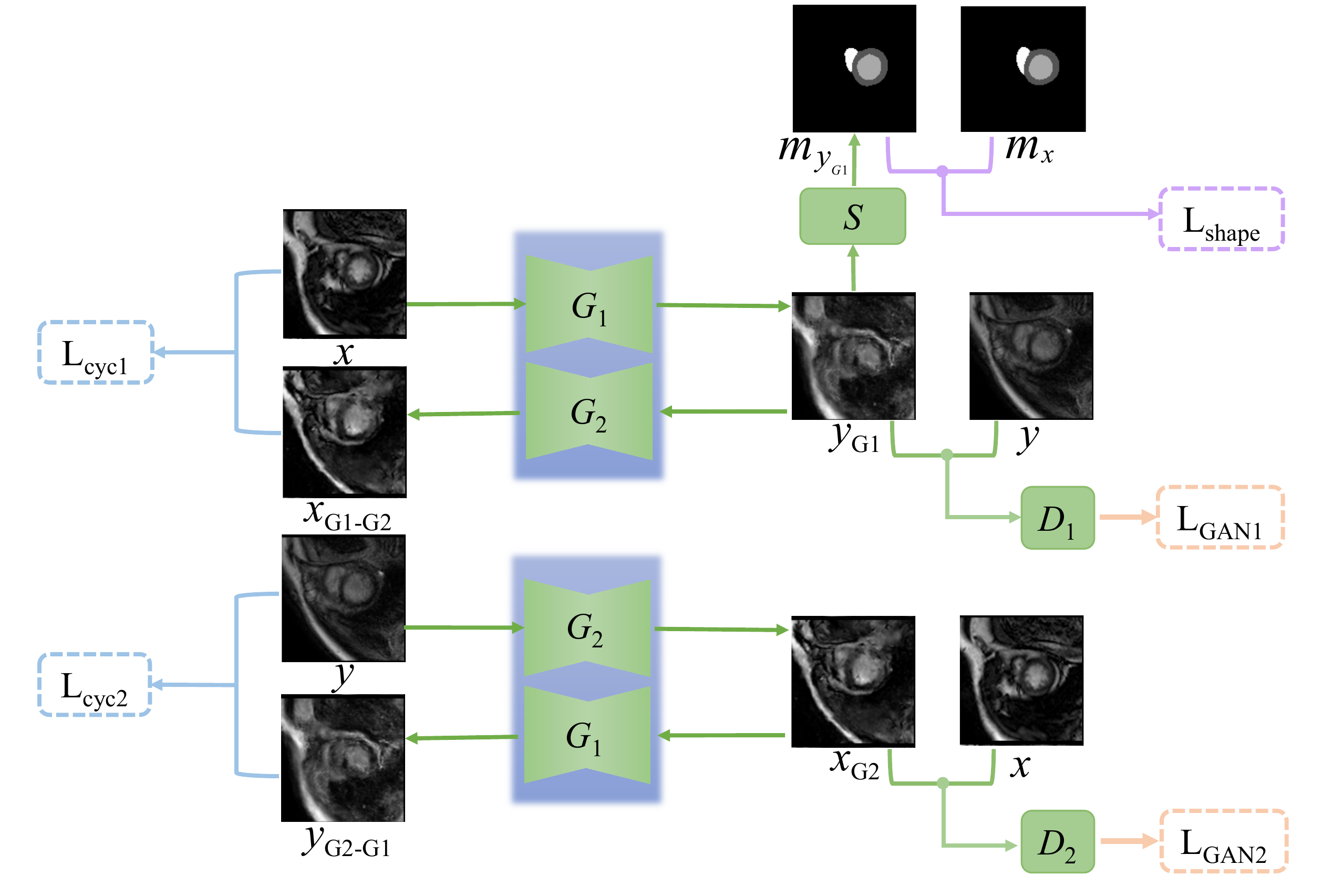}
	\caption{Building block of shape-transfer GAN, whose loss includes three parts: adversarial learning loss (L$_{gan}$), cycle-reconstruction loss (L$ _{cyc} $) and shape preservation loss (L$ _{shape} $). G, D and S represent generator, discriminator and segmentation network respectively. And $x$, $y$ represent bSSFP and LGE images respectively.} 
	\label{fig1}
\end{figure}
The proposed Shape-Transfer GAN can learn a mapping functions between two domains bSSFP and LGE, with the anatomical shape of myocardium in the bSSFP preserved while the intensity distribution being changed into the style of LGE image. To obtain the myocardium shape and enforce the shape preservation loss, a segmentation module is also embedded in the generator. Once the adversarial learning is completed, the segmentation module can be directly applied to novel LGE images for myocardium segmentation. Fig.~\ref{fig1} gives the building block of shape-transfer GAN, which contains three blocks: 1) adversarial learning (L$_{gan}$), where two generators and two discriminators are learned to generate realistic LGE images from bSSFP images, and also the inverse mapping; 2) Cycle-reconstruction learning (L$ _{cyc} $), where the quality of the generated images are improved by the constraint of re-generating the original input image~\cite{ref_article5}; and 3) shape-preservation learning (L$ _{shape} $), where the gen-erated LGE images are constrained to preserve the anatomic shape of the input bSSFP image, and a segmentation model is embedded in the generator and learned in the meantime.
\subsection{Adversarial Learning}
We introduce two generators G$ _{1} $, G$ _{2} $, and two adversarial discriminators D$ _{1} $ and D$ _{2} $ ,where D$ _{1} $ aims to distinguish between real LGE images $\left\lbrace y\right\rbrace $ and the generated ones by $ \left\lbrace G_{1}\left( x\right) \right\rbrace  $ from bSSFP images and D$ _{2} $ to distinguish between real bSSFP images $\left\lbrace x\right\rbrace $ the and generated ones by $ \left\lbrace G_{2}\left( y\right) \right\rbrace  $ from LGE images. In such a way, a bidirectional mapping function can be learned for the two image domains. The objective function of adversarial learning is:
\begin{equation}
 L_{GAN1} = \mathbb{E}_{y\sim P_{LGE(y)}}[\log D_{1}(y)] + \mathbb{E}_{x\sim P_{bSSFP(x)}}[\log (1-D_{1}(G_{1}(x)))]
\end{equation}
\begin{equation}
L_{GAN2} =  \mathbb{E}_{x\sim P_{bSSFP(x)}}[\log D_{2}(x)] + \mathbb{E}_{y\sim P_{LGE(y)}}[\log (1-D_{2}(G_{2}(x)))]
\end{equation}
\begin{equation}
L_{GAN}(G_{1},G_{2},D_{1},D_{2}) = \frac{1}{2}(L_{GAN1} + L_{GAN2})
\end{equation}
where $ P_{bSSFP}\left( x\right)  $ and $ P_{LGE}\left( y\right) $ are the data distributions of the bSSFP and LGE images, respectively.

\subsection{Cycle-reconstruction Learning}
To ensure meaningful information can be well kept during the domain mapping of the adversarial learning procedure, we introduce the cycle-reconstruction learning block. Only the previous generator and discriminator cannot necessarily lead to a good domain mapping, due to the oscillation learning procedure. The discriminator only makes global image-level decision of whether an image is fake or real, while the detailed local information cannot be guaranteed. Given this consideration, the cycle-reconstruction learning block is introduced, which re-generated the original image of source domain from the generated images in the target domain. A good mapping should keep well structure information of the source domain during this cycle-reconstruction procedure. We express the objective of cycle-reconstruction learning as:
\begin{equation}
L_{cyc1} = \mathbb{E}_{x\sim P_{bSSFP(x)}}[\Vert G_{2}(G_{1}(x)) - x \Vert_{1}]
\end{equation}
\begin{equation}
L_{cyc2} = \mathbb{E}_{y\sim P_{LGE(y)}}[\Vert G_{1}(G_{2}(y)) - y \Vert_{1}]
\end{equation}
\begin{equation}
L_{cyc}(G_{1},G_{2}) = \frac{1}{2}(L_{cyc1} + L_{cyc2})
\end{equation}
\subsection{Shape Preservation Learning}
To make sure the generated LGE images $ \left\lbrace y_{G1} \right\rbrace  $ have clear and correct boundary, we make use of the available myocardium shape masks $ \left\lbrace m_{x} \right\rbrace  $ of the bSSFP images and introduce the shape preservation learning block, where the myocardium shape of the generated fake LGE image is constraint to be identical to that of the input bSSFP image. To achieve this, a segmentation network \textbf{S} is embedded into the generator $ G_{1} $ to obtain the myocardium shape of the generated images. Shape preservation is described by the cross-entropy (CE) loss between the shape $ \textbf{m}_{\textbf{x}} $ of the of real bSSFP image and the output of the segmentation network:
\begin{equation}
L_{shape}(S,G_1) = \mathbb{E}_{x\sim P_{bSSFP(x)}}[CE(m_x,S(G_1(x)))]
\end{equation}
\subsection{Overall Objective}
The overall objective of our shape-transfer GAN is:
\begin{equation}
L_{total}(G_{1},G_{2},D_{1},D_{2},S) = L_{GAN} + \lambda_1L_{cyc} + \lambda_2L_{shape}
\end{equation}
where $ \lambda_{1} $ and $ \lambda_{2} $ adjust the balance of the three terms. After the shape-transfer GAN is learned, the segmentation network S can be directly applied to any novel LGE images.  

\section{Experiment}
We validate our method with the dataset provided by the MS-CMRSeg 2019 challenge. In this section, we first describe the experiment configurations, which include details of the dataset, our experimental setup and the evaluation criterion. Then we report the performance of our method and compare it with existing state-of-art methods. 
\subsection{Experimental Configuration}
\begin{figure}[!b]
	\centering
	\includegraphics[width=0.85\textwidth]{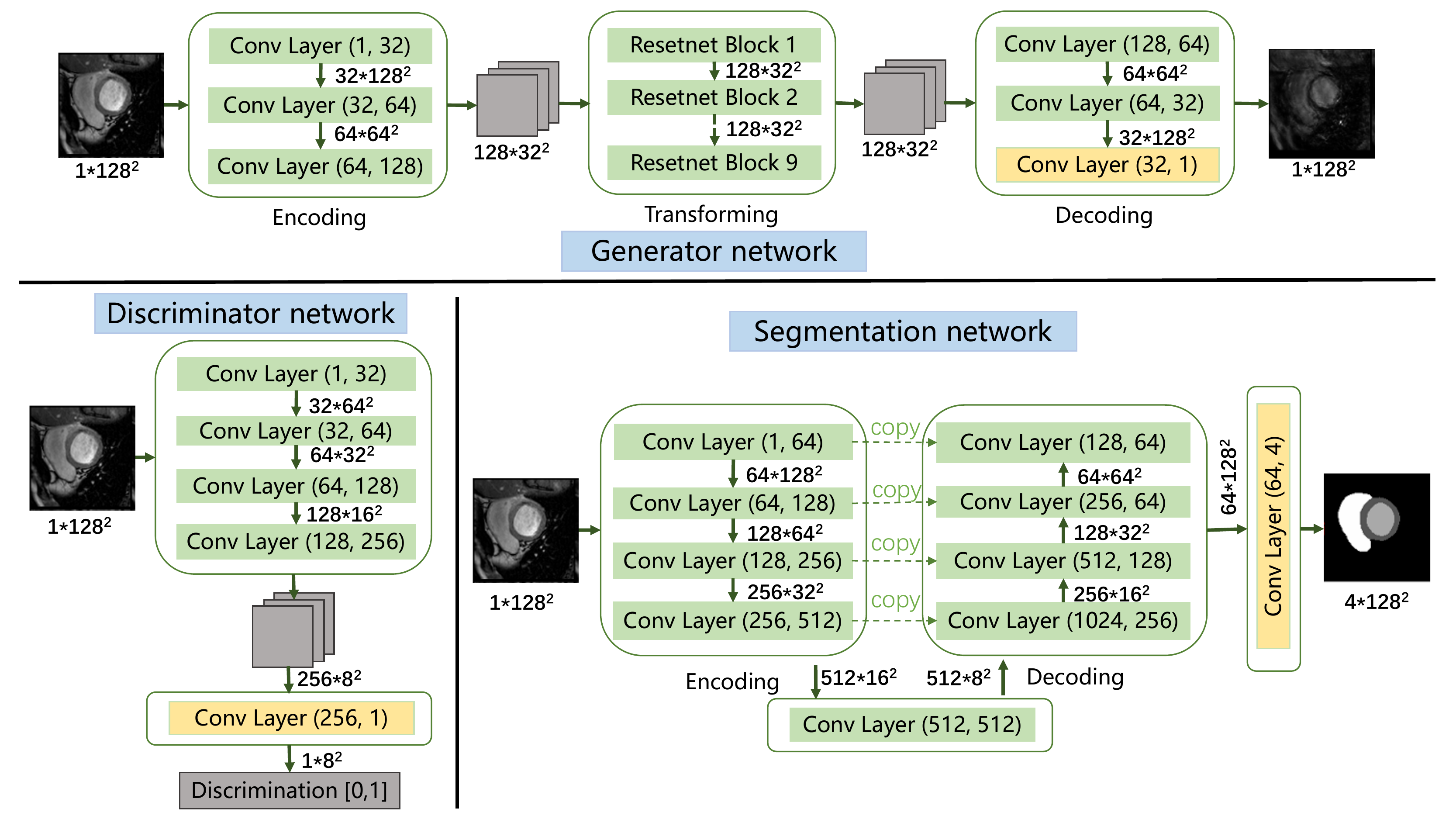}
	\caption{The network details of generator, discriminator and segmentor.} 
	\label{fig2}
\end{figure}
\subsubsection{Dataset.}The Multimodal CMR data (includes bSSFP, LGE and T2 images) used in the paper were collected from 45 patients, where ground truth (GT) of myocardium (Myo), left ventricle (LV) and right ventricle (RV) in 35 patients were provided for bSSFP and T2 images, while for 5 patients GT of LGE images were provided for validation. The rest 40 patients are used for test. For each patient, the bSSFP images consist of 8 - 12 slices, with in-plane resolution of 1.25$ \times $1.25 mm and slice thickness of 8 to 13 mm. The T2 images have 3 - 7 slices, with in-plane resolution of 1.35$ \times $1.35 mm and slice thickness of 12 to 20 mm. The LGE images have 10 - 18 slices with in-plane resolution of 0.75$ \times $0.75 mm and slice thickness of 5 mm. The size of the images range from 256$ \times $256 to 512$ \times $512 and were resized and crop to 128$ \times $128 for Shape-Transfer GAN.

\subsubsection{Experiment setup.} Fig.~\ref{fig2} shows the network details. We used AdamOptimizer with learning rate of 1e-4 for Shape-Transfer GAN and 1e-5 for segmentation network. The input of Shape-Transfer GAN were 2D slices from bSSFP images of 35 patients and LGE images of 45 patients. Note that the segmentation network was pretrained with bSSFP image-label pairs and then the Shape-Transfer GAN was trained for 200 epochs. %The codes and models were implemented using Pytorch.

\subsubsection{Evaluation Metrics.}To evaluate the segmentation performance, Dice score, Jaccard score, average surface distance (ASD) and Hausdorff Distance (HD) were used. Let V$ _{Seg} $ and V$ _{GT} $ be the segmentation and the ground truth volume, and B$ _{Seg} $, B$ _{GT} $ their boundaries. They are computed as:

%The Dice score is defined related to the overlap of the two volumes, i.e.
\begin{equation}
Dice(V_{Seg},V_{GT}) = \frac{2\vert V_{Seg} \cap V_{GT} \vert}{\vert V_{Seg} + V_{GT} \vert},~~
Jaccard(V_{Seg},V_{GT}) = \frac{2\vert V_{Seg} \cap V_{GT} \vert}{\vert V_{Seg} \cup V_{GT} \vert}
\end{equation}

%ASD measures the average distances from points on the boundary of Seg to GT, and d(,) indicate the distance between two points,
\begin{equation}
ASD(B_{Seg},B_{GT}) = \frac{1}{\vert B_{Seg} \vert + \vert B_{GT} \vert} \times \left( \sum\limits_{p\in B_{Seg}}d\left( p,B_{GT}\right)  + \sum\limits_{q\in B_{GT}}d\left( q,B_{Seg}\right) \right) 
\end{equation}
%HD measures how far two subsets of a metric space are from each other, which is defined by,
\begin{equation}
HD(B_{Seg},B_{GT}) = \max \left\lbrace \sup\limits_{p\in B_{Seg}}\inf\limits_{q\in B_{GT}}d\left( p,q\right) ,\sup\limits_{p\in B_{GT}}\inf\limits_{q\in B_{Seg}}d\left( p,q\right) \right\rbrace 
\end{equation}

\subsection{Performance Evaluation and Analysis}
\begin{figure}[!b]
	\centering
	\includegraphics[width=0.85\textwidth]{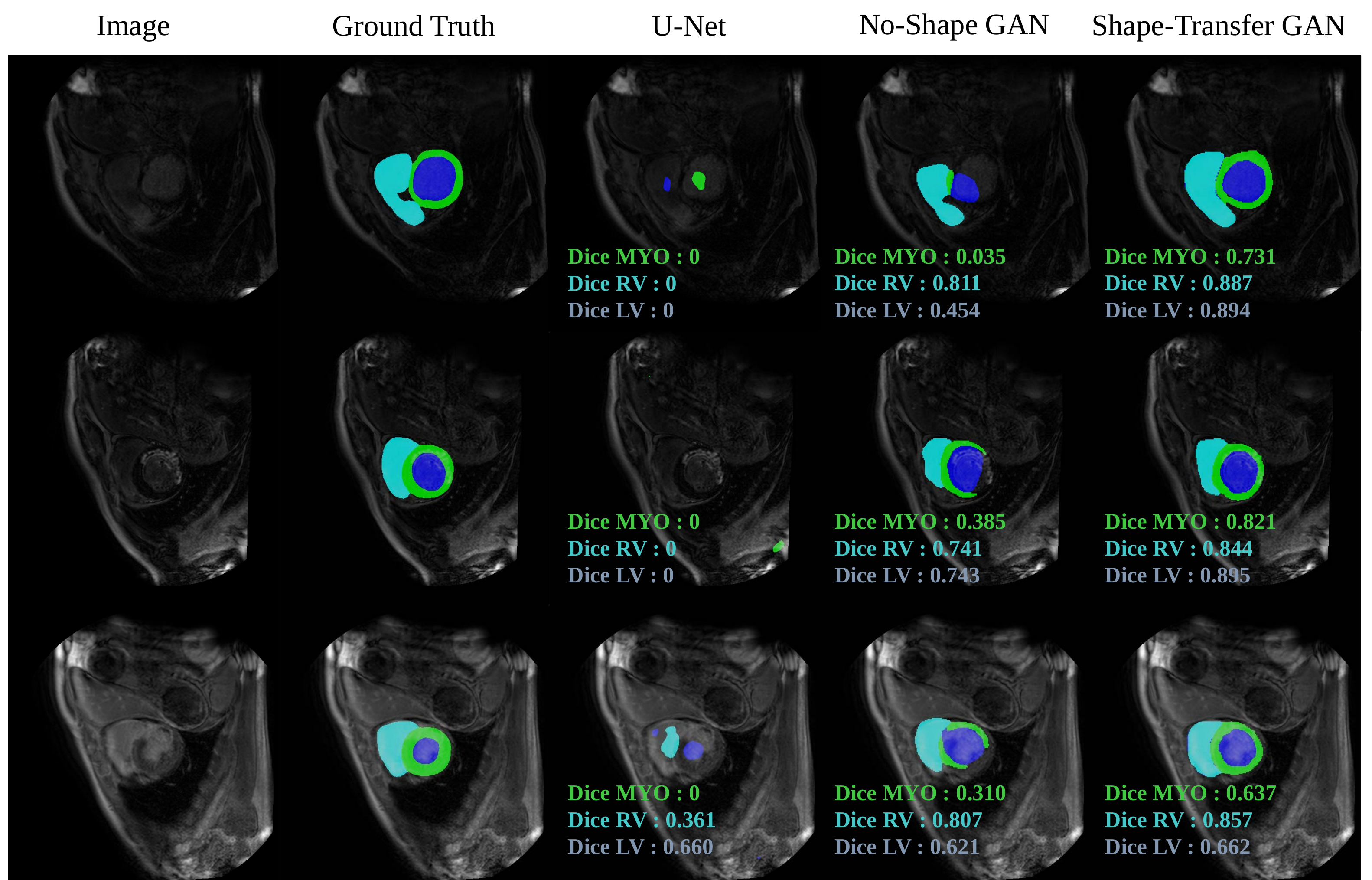}
	\caption{Segmentation results of different methods for ablation study. Each row represents different slice from LGE images. The proposed Shape-Transfer GAN gives the best segmentation results.} 
	\label{fig3}
\end{figure}
\subsubsection{Ablation study.} We first conduct ablation study and validate the effectiveness of our shape-transfer GAN using the LGE images of the 5 patients for validation. The proposed shape-transfer GAN was compared with U-net and GAN with no shape preservation (no-shape GAN). We train the U-net directly with bSSFP images or the generated LGE images, and the provided labels in bSSFP domain. 
%for segmentation of bSSFP images. For no-shape GAN, we first train the generator to obtain fake LGE images and then train the U-net with those fake LGE images and the labels of bSSFP images. To ensure a fair comparison, no label of LGE images are employed during the training procedure. 

As can be drawn from Table~\ref{tab1}, when no adversarial learning is employed, U-net cannot be applied directly to LGE images due to the different intensity distributions. For no-shape GAN, the adversarial learning transfers this distribution from the bSSFP domain to the target LGE domain, therefore make the segmentation network trained with labels of bSSFP domain ready for the LGE domain. But the performance is still far from satisfaction. With the proposed shape preservation learning block, the performance can be clearly improved. Shape-Transfer GAN can keep the myocardium shape accurately in the generated LGE images, thus leads to better synthetic image-label pairs for learning.
 
Fig.~\ref{fig3} shows the visualization results from three different slices for these methods. As can be obviously drawn, when no LGE labels were used for training, U-net cannot capture the shape of myocardium at all. It even makes false positive regions in distant background regions. With adversarial learning, No-shape GAN can well capture the shape of LV, RV and myocardium. However, there are still some regions that are not captured or boundaries that are not well aligned. With the shape-preservation learning, Shape-Transfer GAN can deliver accurate segmentation results. An interest observation from the first row is that a part of RV is missing in the ground truth label, while our method can fill it.

Table~\ref{tab2} shows the performance of our method on the test dataset, which has LGE images of 40 patients (three failure cases were excluded). Without true label information for model training, our method is still capable of segmentation well the LV, RV and myocardium of LEG images. Especially, our method achieves for LV segmentation Dice score of 0.847, ASD of 3.110mm, HD of 17.986mm.

\begin{table}[!t]
	\caption{ Ablation study of our method on validation dataset of 5 patients LGE images. Dice Score $ \left( Mean \pm std \right) $ is presented.} %With U-net as a baseline model and no label of LGE image for training, the adversarial learning and the shape-preservation learning can effectively improve the segmentation performance of LV, RV, and myocardium.
	\label{tab1}
	\setlength{\tabcolsep}{4mm}{
	\begin{tabular}{cccc}
		\toprule
		%&\multicolumn{3}{c}{ Dice Score $ \left( Mean \pm std \right) $ }\\
		Method &  LV & RV & Myo\\
		\midrule
		U-Net &  {0.249 $ \pm $ 0.197} & {0.286 $ \pm $ 0.069} & {0.043 $ \pm $ 0.035}\\
		No-Shape GAN & {0.589 $ \pm $ 0.190} & {0.638 $ \pm $ 0.092} &{0.303 $ \pm $ 0.190}\\
		Shape-Transfer GAN & {0.764 $ \pm $ 0.125} & {0.738 $ \pm $ 0.090} &{0.607 $ \pm $ 0.117}\\	
		\bottomrule
	\end{tabular}}
	\end{table}

\begin{table}[!t]
	\caption{Segmentation performance of Shape-Transfer GAN on test dataset of LGE images from 40 patients with three failure cases excluded.}
	\label{tab2}
	\setlength{\tabcolsep}{5.3mm}{
		\begin{tabular}{cccc}
			\toprule
		%	&\multicolumn{3}{c}{ Dice Score $ \left( Mean \pm std \right) $ }\\
			Metrics &  LV & RV & Myo\\
			\midrule
			Dice &  {0.847 $ \pm $ 0.054} & {0.776 $ \pm $ 0.048} & {0.686 $ \pm $ 0.078}\\
			Jaccard &  {0.738 $ \pm $ 0.079} & {0.636 $ \pm $ 0.063} & {0.527 $ \pm $ 0.087}\\
			\midrule
			Metrics &  LV endo & LV epi & RV endo\\
			\midrule
			ASD(mm) & {3.110 $ \pm $ 1.039} & {3.022 $ \pm $ 0.736} &{3.953 $ \pm $ 0.908}\\
			HD(mm) & {17.986 $ \pm $ 4.028} & {17.453$ \pm $ 5.902} &{21.974$ \pm $ 10.026}\\	
			\bottomrule
	\end{tabular}}
\end{table}

\begin{table}[!t]
	\caption{Performance Comparison of our method and existing state-of-art methods on the same dataset. }
	\label{tab3}
	\setlength{\tabcolsep}{1.7mm}{
		\begin{tabular}{ccccc}
			\toprule
			Dice score&  Shape-Transfer GAN & GMM$ + $bSSFP & MvMM & SRSCN \\
			\midrule
			LV &  {0.847 $ \pm $ 0.054} & {0.836  $ \pm $ 0.071} & {0.866 $ \pm $ 0.063} & {0.915  $ \pm $ 0.052}\\
			RV & {0.776 $ \pm $ 0.048} & - & - &{0.882  $ \pm $ 0.084}\\
			Myo & {0.686 $ \pm $ 0.078} & {0.635 $ \pm $ 0.120} & {0.717 $ \pm $ 0.076} &{0.812 $ \pm $ 0.105}\\	
			\bottomrule
	\end{tabular}}
\end{table}

%(1) the GMM segmentation initialized from the result of Atlas+bSSFP, referred to as GMM+bSSFP~\cite{ref_article2}.

%(2) the MvMM with both FFD and SC registration correction, referred to as MvMM~\cite{ref_article2}.

\subsubsection{Performance comparison.}
Table~\ref{tab3} compares our method with existing state-of-art methods, including two GMM-based methods (GMM+bSSFP, MvGMM) ~\cite{ref_article2}, and one deep neural network based method (SRSCN) ~\cite{ref_article1}. When compared with the GMM-based methods, our method can deliver comparable performance, but with less application complexity. The iterative optimization procedure adds the complexity of the GMM-based methods during practice application. When compared with SRSCN, our method fails to show better or comparable performance. This is due to the fact that SRSCN was trained with ground truth labels of 25 patients' LGE images .

\section{Conclusion}
We propose the Shape-Transfer GAN for cardiac segmentation of LGE MRI images, which can learn the procedure of generating realistic LGE images with the anatomical shape information well kept, and thus obtain an LGE segmentation network. Our method avoided the use of LGE label during the learning of the segmentation. We validated the effectiveness of the proposed shape-transfer technique and tested the final performance on a dataset of 40 patients. The good segmentation results prove that our method has a great potential in cases of medical image segmentation tasks with insufficient labeled data. 
\section{Acknowledgement}
The paper is partially supported by the Natural Science Foundation of China under Grants 61801296, the Overseas High-Caliber Personnel Peacock Plan of Shenzhen, and the start-up funding of Shenzhen University.
% the environments 'definition', 'lemma', 'proposition', 'corollary',
% 'remark', and 'example' are defined in the LLNCS documentclass as well

%
% ---- Bibliography ----
%
% BibTeX users should specify bibliography style 'splncs04'.
% References will then be sorted and formatted in the correct style.
%
% \bibliographystyle{splncs04}
% \bibliography{mybibliography}
%

\end{document}